\documentclass[12pt]{article}
\usepackage{epsfig}
\usepackage{amsfonts}
\begin{document}
\begin{titlepage}
\begin{center}

{\Large Superstatistics}

\vspace{2cm}

{\bf C. Beck}

\vspace{0.5cm}

School of Mathematical Sciences, Queen Mary, University of London,
Mile End Road, London E1 4NS, UK

\vspace{1cm} {\bf E.G.D. Cohen}

\vspace{0.5cm}

The Rockefeller University, 1230 York Avenue, New York, NY
10021-6399

\vspace{3cm}

\end{center}

\abstract{We consider nonequilibrium systems with complex dynamics
in stationary states with large fluctuations of intensive quantities
(e.g. the temperature, chemical potential or energy dissipation) on long
time scales.
Depending on the statistical properties of the
fluctuations, we obtain different effective statistical mechanical
descriptions. Tsallis statistics follows from a $\chi^2$-distribution
of an intensive variable, but other classes of
generalized statistics are obtained as well. We show that for
small variance of the fluctuations all these different statistics
behave in a universal way.}

\vspace{1.3cm}

\end{titlepage}

Einstein never accepted Boltzmann's principle $S=k \log W$,
because he argued that the statistics ($W$) of a system should
follow from its dynamics and, in principle, could not be
postulated {\em  a priori} \cite{pais, cohen}. Remarkably, the
Boltzmann-Gibbs (BG) statistics works perfectly for classical
systems with short range forces and relatively simple dynamics in
equilibrium. A possible consequence of Einstein's criticism is
that for systems with sufficiently complex dynamics, other than BG
statistics might be appropriate. Such a statistics has indeed been
proposed by Tsallis \cite{tsa1} and has been observed in various
complex systems \cite{wilk}--\cite{cosmic}.
For these types of systems the formalism of nonextensive statistical mechanics 
\cite{tsa2,abe} is a useful theoretical concept,
in the sense that a more general
Boltzmann factor is introduced which depends on an entropic index
$q$ and which, for $q=1$, reduces to the ordinary Boltzmann factor. 
Applications of the formalism have been mainly reported
for multifractal systems, systems with long-range interactions and
nonequilibrium systems with a stationary state. Several types of
generalized stochastic dynamics have been recently constructed for
which Tsallis statistics can be proved rigorously
\cite{wilk,prl,jap}. Physical applications 
include 2-d and 3-d turbulence \cite{baraud}--\cite{ari}, momentum
spectra of hadronic particles produced in $e^+e^-$ annihilation
experiments \cite{bedi,physa}, the statistics of cosmic rays
\cite{cosmic}, and many other phenomena.

In this letter we will quite generally deal with nonequilibrium
systems with a long-term stationary state that possess a
spatio-temporally fluctuating intensive quantity. We will show
that after averaging over the fluctuations one can obtain as an
effective description not only Tsallis statistics but an infinite
set of more general statistics, which we will call
`superstatistics'. We will consider several concrete examples. A
parameter $q$ can be defined for all these new statistics and
given a physical interpretation. For
$q= 1$ all superstatistics reduce to the Boltzmann factor $e^{-\beta_0 E}$
of ordinary statistical mechanics. One of our main results is
universality for small amplitudes of the fluctuations. If the
variance of the fluctuations is small, then the first-order
corrections to the ordinary Boltzmann factor are the same for all
superstatistics (i.e.\ those of Tsallis statistics). On the other
hand, for large variances quite different behaviour from Tsallis
statistics can be generated. We calculate the higher-order
correction terms of various examples of superstatistics.

Experimental evidence for superstatistics that is nearly Tsallis
but has tiny corrections beyond Tsallis has
very recently been provided by Jung and Swinney \cite{jung}, by
careful analysis of velocity fluctuations
in a turbulent Taylor-Couette flow. Moreover,
experimental studies from fusion plasma physics provide
possible evidence for
non-Tsallis superstatistics based on F-distributions \cite{sattin}.
All this illustrates the need for a more general approach towards generalized
statistics for complex dynamical systems.

Let us now outline the theory.
Consider a driven nonequilibrium system that is composed of regions that
exhibit spatio-temporal fluctuations of an intensive quantity.
This could be the inverse temperature $\beta$, or equally well the
pressure, chemical potential, or the energy dissipation rate in a
turbulent fluid \cite{prl,reynolds,pla}. We will select the
temperature here as our fluctuating quantity but it could be
anything of the above.
We note that even the fluctuation theory of Onsager and Machlup
does not treat this case and is restricted to fluctuations
of extensive thermodynamic variables \cite{onsager}.
We consider a nonequilibrium steady state of
macroscopic system, 
made up of many smaller cells that are
temporarily in local
equilibrium. Within each cell,
$\beta$ is approximately constant.
Each cell is large enough to obey statistical mechanics,
but has a different $\beta$ assigned to it, according to a
probability density $f(\beta)$. We assume that the local
temperatures in the various cells change on a long time scale $T$.
This time scale $T$ is much larger than the relaxation time that
the single cells need to reach local equilibrium.
It is clear that
our model is a suitable approximation for a continuously varying
temperature field that has spatial correlation length of the order
of the cell size $L$ and
a temporal correlation length $T$. 

A Brownian test particle moves
for a while in a certain cell with a given temperature, then moves
to the next cell, and so on. Its velocity $v$ obeys
\begin{equation}
\dot{v}=-\gamma v+\hat{\sigma} L(t),
\end{equation}
where $L(t)$ is Gaussian white noise. The inverse temperature of
each cell is related to the parameters $\gamma$ and
$\hat{\sigma}$ by $\beta=\gamma/\hat{\sigma}^2$. However, unlike
ordinary Brownian motion, the parameter $\beta$ is not constant but
changes temporally on the time scale $T$ and spatially
on the scale $L$. These changes are ultimately produced by
the very complex dynamics of the environment
of the Brownian particle.
It has been proved
by one of us \cite{prl} that after averaging over the
fluctuating $\beta$ this simple
generalized Langevin model generates Tsallis statistics for
$v$ if $\beta$ is a $\chi^2$-distributed random variable. Moreover,
the obtained distributions for $v$ were shown to fit quite precisely 
distributions of longitudinal velocity differences as measured
in turbulent Taylor-Couette flows \cite{prl,BLS}, as well as 
measurements in Lagrangian turbulence \cite{pla,boden}.

Let us here generalize this approach to general distributions
$f(\beta)$ and general (effective) Hamiltonians.
In the long-term run $(t>>T)$, the stationary probability density of our
nonequilibrium system arises out of Boltzmann factors $e^{-\beta
E}$ associated with the cells that are averaged over the various fluctuating inverse
temperatures $\beta$. If $E$ is the energy of a microstate associated
with each cell, 
we may write
\begin{equation}
B (E) =\int_0^\infty d\beta f(\beta) e^{-\beta E}, \label{gb}
\end{equation}
where $B$ is a kind of effective Boltzmann factor for our nonequilibrium
system, the superstatistics of the system. 
In a sense,
it represents the statistics of the statistics ($e^{-\beta E}$) of the cells
of the system.
$B(E)$ may significantly
differ from the ordinary Boltzmann factor, which is recovered for
$f(\beta)=\delta (\beta-\beta_0)$. 
For the above simple example of a Brownian test particle
of mass $1$ one has $E=\frac{1}{2}v^2$, so that the
long-term stationary state consists of a superposition of Gaussian distributions
$e^{-\beta E}$ that are weighted with the probability density $f(\beta)$
to observe a certain $\beta$. But our consideration in
the following applies to arbitrary energies $E$ associated with
the cells, not only $E=\frac{1}{2}v^2$. The central hypothesis
of our paper is that generalized Boltzmann factors of the form (\ref{gb})
are physically relevant for large classes of dynamically
complex systems with fluctuations.

One immediately recognizes that the generalized Boltzmann factor
of superstatistics is given by the Laplace transform of the
probability density $f(\beta)$. Although there are infinitely many
possibilities, certain criteria must be fulfilled which
significantly reduce the number of physically relevant cases:

\begin{enumerate}
\item $f(\beta)$ cannot be any function but must be a normalized
probability density. It may, in fact, be a physically relevant
density from statistics, say Gaussian, uniform, chi-squared,
lognormal, but also other, as yet unidentified, distributions 
could be considered if the underlying dynamics is
sufficiently complex.

\item The new statistics must be normalizable, i.e.
the integral $\int_0^\infty B(E)dE$ must exist, or in general
the integral $\int_0^\infty \rho (E) B(E)dE$, where $\rho (E)$
is the density of states.

\item The new statistics should reduce to BG-statistics
if there are no fluctuations of intensive quantities at all.

\end{enumerate}

We will now consider several examples of superstatistics (later we
will see what is universal to all of them). The distribution $f(\beta)$
is ultimately determined by the invariant measure
of the underlying dynamical system and {\em a priori}
unknown. Gaussian $\beta$-fluctuations
would generate
one of the simplest $f(\beta)$, but these types of models
are unphysical since they allow for negative $\beta$ with some
non-zero probability. Instead, we 
need
distributions where the random variable $\beta$ is always positive.


\subsection*{Uniform distribution}

As a very simple model where everything can be calculated
analytically let us first consider a uniform distribution
of $\beta$ on some interval $[a,a+b]$, i.e.
\begin{equation} f(\beta) =\frac{1}{b}
\end{equation}
for $0\leq a\leq \beta \leq a+b$, whereas $f(\beta)=0$ elsewhere. The
mean of this distribution is
\begin{equation}
\beta_0=a+\frac{b}{2}
\end{equation}
and for the variance one obtains $\sigma^2=\langle \beta^2\rangle
-\beta_0^2=b^2/12$. The superstatistics of this model follows from
the generalized Boltzmann factor
\begin{equation}
B=\int_0^\infty e^{-\beta E}f(\beta)d\beta =  \frac{1}{bE} \left(
e^{-(\beta_0-\frac{1}{2}b)E}-e^{-(\beta_0+\frac{1}{2}b)E} \right) .
\end{equation}
This is normalizable for any $E\geq 0$. Hence this is a reasonable
model of superstatistics. For small $bE$ one obtains, by series
expansion of $e^{-\frac{1}{2}bE}$,
\begin{equation}
B=e^{-\beta_0E}(1+\frac{1}{24}b^2E^2+\frac{1}{1920}b^4E^4+...)
\label{2}
\end{equation}
Clearly, for $b\to 0$ ordinary statistical mechanics is recovered.

\subsection*{2-level distribution}

Suppose the subsystems can switch between two different discrete
values of the temperature, each with equal probability. A physical
example might be a Brownian particle that can switch between two
different states with different friction constants, or two different chemical
potentials. The probability density is given by
\begin{equation} f(\beta) =\frac{1}{2}\delta
(a)+\frac{1}{2} \delta (a+b).
\end{equation}
The average of $\beta$ is again given by
\begin{equation}
\beta_0=a+\frac{b}{2}
\end{equation}
and for the variance one obtains $\sigma^2=\langle \beta^2\rangle
-\beta_0^2=b^2/4$. The generalized Boltzmann factor becomes
\begin{equation}
B=\int_0^\infty e^{-\beta E}f(\beta)d\beta = e^{-\beta_0 E}
\frac{1}{2} \left( e^{\frac{1}{2}bE}+e^{-\frac{1}{2}bE} \right) .
\end{equation}
This is normalizable for any $E\geq 0$. For small $bE$ one obtains
\begin{equation}
B=e^{-\beta_0E}(1+\frac{1}{8}b^2E^2+\frac{1}{384}b^4E^4+...) .
\label{3}
\end{equation}

\subsection*{Gamma-distribution}

The assumption of a Gamma (or $\chi^2$-) distributed inverse
temperature $\beta$ leads to Tsallis statistics, so far the most
relevant example of superstatistics. We may write the Gamma
distribution as
\begin{equation}
f (\beta) = \frac{1}{b \Gamma (c)} \left(
\frac{\beta}{b}\right)^{c-1} e^{-\frac{\beta}{b}},  \label{flucc}
\end{equation}
where $c>0$ and $b>0$ are parameters. The number $2c$ can be
interpreted as the effective number of degrees of freedom
contributing to the fluctuating $\beta$. The average of $\beta$ is
$\beta_0 =\int_0^\infty \beta f(\beta) d\beta =bc$ and for the
variance of $\beta$ one obtains $\sigma^2=b^2c$.

The Gamma distribution naturally arises if $n=2c$ independent
Gaussian random variables $X_k$ with average 0 are squared and
added. If we write
\begin{equation} \beta
=\sum_{k=1}^n X_k^2
\end{equation}
then this $\beta$ is Gamma distributed with density (\ref{flucc}).
The average of $\beta$ is of course given by $n$ times the
variance of the Gaussian random variables $X_k$. In this sense
the Gamma distribution arises very naturally for a fluctuating
environment
with a finite number $n$ of degrees of freedom.

The integration over $\beta$ yields the generalized Boltzmann
factor
\begin{equation}
B=\int_0^\infty e^{-\beta E} f(\beta ) d\beta =(1+b E )^{-c},
\label{tsallis-d}
\end{equation}
i.e.\ the generalized Boltzmann factor $(1+(q-1)\beta_0
E)^{-\frac{1}{q-1}}$ of nonextensive statistical mechanics
\cite{tsa1,tsa2,abe} if we identify $c=\frac{1}{q-1}$ and
$bc=\beta_0$. It is normalizable for $c>1$ (if $\rho (E)=1$).

We may also write $ B= \exp \left\{ -c \log (1+b E) \right\}$ and
expand the logarithm for small $bE$. The result can be written in
the form
\begin{equation}
B=e^{-\beta_0 E} (1+\frac{1}{2} (q-1) \beta_0^2 E^2 -
\frac{1}{3}(q-1)^2\beta_0^3E^3+...) \label{4}
\end{equation}
where $q$ is the entropic index of nonextensive statistical
mechanics.

\subsection*{Log-normal distribution}

The log-normal distribution
\begin{equation}
f(\beta) = \frac{1}{\beta s \sqrt{2\pi}}\exp\left\{ \frac{-(\log
\frac{\beta}{m})^2}{2s^2}\right\}
\end{equation}
yields yet another possible superstatistics (see also \cite{cast}
for a related turbulence model); $m$ and $s$ are parameters. The average
$\beta_0$ of the above log-normal distribution is given by
$\beta_0=m\sqrt{w}$ and the variance by $\sigma^2=m^2w(w-1)$,
where $w:= e^{s^2}$. The generalized Boltzmann factor
$B=\int_0^\infty f(\beta)e^{-\beta E}d\beta$ cannot be evaluated
in closed form, but in leading order we obtain for small variance
of the inverse temperature fluctuations
\begin{equation} B=e^{-\beta_0 E}
(1+\frac{1}{2}m^2w(w-1)E^2-\frac{1}{6}m^3w^\frac{3}{2}
(w^3-3w+2)E^3+...) . \label{5}
\end{equation}

\subsection*{F-distribution}

The last example we want to consider is that of a $\beta\in
[0,\infty]$ distributed according to the F-distribution
\cite{hast}
\begin{equation}
f(\beta) =\frac{\Gamma ((v+w)/2)}{\Gamma (v/2) \Gamma (w/2)}
\left( \frac{bv}{w} \right)^{v/2} \frac{\beta^{\frac{v}{2}-1}}{(1+
\frac{vb}{w}\beta)^{(v+w)/2}} .
\end{equation}
Here $w$ and $v$ are positive integers and $b>0$ is a parameter.
We note that for $v=2$ we obtain a Tsallis distribution. However,
this is
a Tsallis distribution in
$\beta$-space, not in $E$-space as in eq.~(\ref{tsallis-d}).

The average of $\beta$ is given by
\begin{equation}
\beta_0=\frac{w}{b(w-2)}
\end{equation}
and the variance by
\begin{equation}
\sigma^2=\frac{2w^2(v+w-2)}{b^2 v (w-2)^2(w-4)} .
\end{equation}

The Laplace transform cannot be obtained in closed form, but for
small variance of the fluctuations we obtain the series expansion
\begin{eqnarray}
B(E)&=&e^{-\beta_0E} \left(
1+\frac{w^2(v+w-2)}{b^2v(w-2)^2(w-4)}E^2 \right. \nonumber \\ &\,&
- \left.
\frac{4w^3(v+w-2)(2v+w-2)}{3b^3v^{3/2}(w-2)^3(w-4)(w-6)}E^3+...\right) .
\label{6}
\end{eqnarray}
F-distribution in $\beta$-space have also been recently studied by
Sattin and Salasnich \cite{sattin}, who point out
possible applications in fusion plasma physics.

While in general the large-$E$ behavior is different for all superstatistics
(it strongly depends on the function $f(\beta)$), we now show that
the low-$E$ behavior is universal.
We note that for the above five
distribution functions the first-order corrections to the Boltzmann
factor $e^{-\beta_0E}$ in eqs. (\ref{2}), (\ref{3}), (\ref{4}), (\ref{5}),
(\ref{6}) can all be written in a universal form. 
When expressing the generalized Boltzmann
factor in terms of the average temperature $\beta_0$ and the
variance $\sigma^2$ of the various probability densities $f (\beta
)$, one always obtains the same result for small $\sigma$,
\begin{equation}
B=e^{-\beta_0 E} (1+\frac{1}{2}\sigma^2 E^2+O(\sigma^3E^3)),
\label{7}
\end{equation}
where $\sigma$ is the standard deviation of the distribution
$f(\beta)$. Thus in the limit of approaching ordinary statistics,
the generalized Boltzmann factor of superstatistics has a
universal quadratic correction term $(1+\frac{1}{2}\sigma^2 E^2)$.

Let us now prove this universality. For any
distribution $f(\beta)$ with average $\beta_0 :=\langle \beta
\rangle$ and variance $\sigma^2:=\langle \beta^2 \rangle
-\beta_0^2$ we can write
\begin{eqnarray}
B&=& \langle e^{-\beta E} \rangle = e^{-\beta_0 E} e^{+\beta_0 E}
\langle e^{-\beta E}\rangle   \nonumber
\\ &=& e^{-\beta_0 E} \langle e^{-(\beta -\beta_0)E}\rangle
\nonumber \\ &=& e^{-\beta_0 E}
\left(1 +\frac{1}{2}\sigma^2 E^2 +\sum_{r=3}^\infty
\frac{(-1)^r}{r!} \langle (\beta-\beta_0)^r \rangle E^r \right).
\end{eqnarray}
Here the coefficients of the powers $E^r$ are the $r$-th moments
of the distribution $f(\beta)$ about the mean, which can be
expressed in terms of the ordinary moments as
\begin{equation}
\langle (\beta-\beta_0)^r \rangle =\sum_{j=0}^r \left(
\begin{array}{c} r \\ j \end{array} \right)
\langle \beta^j \rangle (-\beta_0)^{r-j} .
\end{equation}

Due to the universality proved above, it now makes sense
to define a universal parameter $q$ for {\em any} superstatistics,
not only for Tsallis statistics.
By comparing
eq.~(\ref{4}) and (\ref{7}), we generally define 
a parameter $q$ by
the relation
\begin{equation}
(q-1)\beta_0^2=\sigma^2,
\end{equation}
or equivalently
\begin{equation}
q=\frac{\langle \beta^2 \rangle}{\langle \beta \rangle^2}.
\label{30}
\end{equation}
For example, for the log-normal distribution we obtain from
eq.~(\ref{30}) $q=w$ and for the F-distribution
$q=1+\frac{2(v+w-2)}{v(w-4)}$. The physical meaning of this
generally defined parameter $q$ is that
$\sqrt{q-1}=\frac{\sigma}{\beta_0}$ is just the coefficient of
variation of the distribution $f(\beta)$, defined by the ratio of
standard deviation and mean. If there are no fluctations of
$\beta$ at all, we obtain $q=1$ as required. For small $\sigma E$
the corrections to the ordinary Boltzmann factor $e^{-\beta_0 E}$
are universal and given by the factor $(1+(q-1)\beta_0^2E^2/2)$.

Since for small $\sigma E$ all superstatistics are the same
and given by Tsallis statistics,
we also have a maximum entropy principle 
for the normalized Boltzmann
factors of all our superstatistics in terms of the Tsallis
entropies \cite{tsa1,tsa2} with an entropic index $q$ given by
eq.~(\ref{30}). For sufficiently
small variance of the fluctuations
they all extremize the Tsallis entropies subject to given constraints, no
matter what the precise form of $f(\beta)$ is. 

However, differences arise if $\sigma E$ is not small. Then there
are many different superstatistics described by different $B(E)$, and
the densities given by $B(E)$ do not extremize the Tsallis
entropies in general. Rather,
as shown very recently by Tsallis and Souza \cite{souza},
they extremize more general classes
of entropy-like functions. The Boltzmann factor of the examples of
superstatistics that we considered above, if expressed in terms of
the universal parameters $q$ and $\beta_0$, can be written as
\begin{equation}
B(E)=e^{-\beta_0E}(1+\frac{1}{2}(q-1)\beta_0^2 E^2+ g(q) \beta_0^3
E^3+...),
\end{equation}
where the function $g(q)$ depends on the superstatistics chosen.
We obtain
\begin{eqnarray}
g(q) &=& 0 \;\;\;\;(uniform \; and \; 2-level) \\
    &=& - \frac{1}{3}(q-1)^2 \;\;\;\;(Gamma) \\
    &=& - \frac{1}{6}(q^3-3q+2) \;\;\;\;(log-normal) \\
    &=&  - \frac{1}{3} \frac{(q-1)(5q-6)}{3-q} \;\;\;\;(F \; with \; v=4) .
\end{eqnarray}
For sufficiently large variance $\sigma^2$ or energy $E$ one can
distinguish between Tsallis
statistics and other superstatistics by the third-order term
$g(q)\beta_0^3 E^3$, which describes the first significant differences
between the various superstatistics and hence provides
more information on 
the underlying complex dynamics. In the turbulence
applications described theoretically in \cite{prl,reynolds,pla}
and experimentally in \cite{BLS, baraud}, the fluctuating
intensive quantity is related to the energy dissipation rate of
the flow, and the fluctuations of this quantity are indeed large.
Moreover, the energy $E$ essentially corresponds to the square of
local velocity differences in the fluid, and these are also measured up to very
large values. Therefore these experiments can well provide
precise information about the kind of superstatistics needed to describe
the complex dynamics.

An interesting question is what type of averaging process is the 
physically most relevant one for a generic system
with intensive parameter fluctuations. We may either work with
unnormalized Boltzmann factors $e^{-\beta E}$ that are 
averaged over $\beta$ and then finally
normalize by doing the integration
over all energy states $E$ (type-A superstatistics), or we may 
work with locally normalized distributions 
$p(E)=\frac{1}{Z(\beta )}e^{-\beta E}$ and average those
over all $\beta$ 
(type-B superstatistics). Since in general the normalization
constant $Z$ depends on $\beta$, the result will differ slightly.
However, case B can be easily reduced to case A
by replacing the distribution $f(\beta)$ by a new distribution
$\tilde{f}(\beta):=CZ(\beta)^{-1}f(\beta)$, where $C$ is a suitable
normalization constant. In other words, type-B superstatistics
with $f$ is equivalent to type-A superstatistics with $\tilde{f}$.
Our formula (\ref{30}) relating $q$ and the variance of the $\beta$ fluctuations
is valid for both type-A and type-B superstatistics, just that
all expectations $\langle ...\rangle$
must be formed with either $f$ (type-A) or $\tilde{f}$
(type-B).
For example, if $E=\frac{1}{2}u^2$, then $Z(\beta)=\sqrt{2\pi /\beta}$
and hence $\tilde{f}(\beta) \sim \beta^{1/2} f(\beta)$.
If $f(\beta)$ is given by the $\Gamma$-distribution eq.~(\ref{flucc}), 
then type-A superstatistics
yields $q=1+\frac{1}{c}$, whereas type-B yields $q=1+\frac{2}{2c+1}$.

To summarize, in this letter we have introduced a class of
generalized statistics, which we have called
`superstatistics'. Tsallis statistics is a special case of
these superstatisics. The
dynamical parameter $q$ of eq.~(\ref{30})
can be defined for all these new
statistics. For small variance of the fluctuations we have shown
that there is universal behavior of all superstatistics. For large
variance there are differences
which provide information on the underlying
complex dynamics. In general, complex nonequilibrium
problems may require different types of superstatistics. Tsallis
statistics is just one example of many
possible new statistics. There is no {\em a priori} reason to
expect that other superstatistics would not be present in nature,
thus confirming the relevance of Einstein's remark.


\end{document}